\documentclass{aa}  

\usepackage{graphicx}
\usepackage{txfonts}
\usepackage{xcolor}
\usepackage{mhchem}
\usepackage{xurl}
\newcommand{\rev}[1]{#1}
\usepackage{hyperref}
\begin{document}

   \title{A neural-network model to find correlations in astrochemical space}
%   \subtitle{A deep learning study on hidden correlations in chemical space}

   \author{D. G. Vargas,
          \inst{1}
          D. Barreiro-Lage,\inst{2}
          M. Agúndez,\inst{2}
          C. Cabezas,\inst{2}
         J. Cernicharo ,\inst{2}
         G. Molpeceres\inst{2}
          }
   \authorrunning{Vargas et al}
   \institute{             
         Universidad Nacional de Colombia, Sede Bogotá, Facultad de Ciencias, Observatorio Astronómico Nacional, Ciudad Universitaria, Bogotá, Colombia
         \and
           Departamento de Astrofísica Molecular, Instituto de Física Fundamental (IFF-CSIC), C/ Serrano 121, 28006 Madrid, Spain \\
              \email{german.molpeceres@iff.csic.es}
             }

   \date{Received \today; accepted \today}

 \abstract{The detection of new interstellar molecules is a central objective of astrochemistry, and arguably the most transformative development in the field over the past decade has been the dramatic increase in both the number and diversity of detected species. Using state-of-the-art natural language processing models, we develop a machine-learning pipeline, in line with recent approaches reported in the literature, to establish relationships between molecular structures, encoded as chemical strings, and molecular column densities in the prototypical dark cloud TMC-1. The workflow generates astrochemically informed embeddings that can be employed in subsequent supervised learning tasks. In this work, these representations are used within a neural-network framework to perform column-density regression on both the original molecular inventory and out-of-distribution species. The resulting model achieves mean squared error values of 0.17/0.58 and mean absolute error values of 0.29/0.59, both expressed in logarithmic units of the column density, together with coefficients of determination of 0.88/0.51 for the train+validation and validation datasets, respectively. The latent vectors learned by the model represent a rich mathematical object that captures not only the underlying chemical relations, but also the potential interstellar detectability of the molecule. Using these vectors in molecules obtained from systematic permutations of the TMC-1 inventory, allow us to identify several promising candidates for future detection in this cloud, together with a large number of apolar species. Furthermore, by training the model on datasets reflecting different historical epochs of molecular detections in TMC-1, we show that recent molecular detections have substantially reshaped our understanding of interstellar chemistry, in a transformative, rather than incremental, way. Beyond accurate prediction of column-density trends for both in-distribution and out-of-distribution molecules, the latent-space representations produced by the model provide a steping stone for developing new forms of data-driven astrochemical intuition and exploration.
 } 
% 5 {} token are mandatory
 
%  \abstract
  % aims heading (mandatory)
 %  {}   

      \keywords{ISM: molecules -- Astrochemistry -- methods: statistical -- methods: data analysis -- ISM: individual objects: TMC-1
                  }

      \maketitle
%
%-------------------------------------------------------------------

\section{Introduction}

One of the main challenges in astrochemistry is to understand the chemical composition of the interstellar medium (ISM) in general, and in some regions that are considered as prototype of a specific chemistry. The chemical inventory of the ISM is continuously expanding, with a particularly strong boost in the last years. Taking as reference the number of molecules reported in the 2018 census of interstellar molecules \citep{mcguire_2018_2018} of 204 molecules, the number of molecules detected in the ISM has increased in more than 70\% (346 at the time of writing). This is a staggering figure considering the evolution of the discipline since the detection of the first molecule, \ce{CH} \citep{1937ApJ....86..483S}. The detection of new interstellar molecules is undoubtedly not incremental, but has had a transformative impact on our understanding of the chemistry of the ISM. For example, the detections in recent years have shown a molecular panorama full of complex organics \citep{Rivilla2020, Rivilla2021, jimenez-serra_precursors_2022, rivilla_first_2023}, exotic cumulenic species \citep{Cernicharo2021c, Cernicharo2021a, 2023A&A...680L...4C, remijan_astronomical_2023, cabezas_discovery_2022,esplugues_discovery_2026}, new metal-bearing compounds \citep{cernicharo_magnesium_2023, gupta_calcium_2024,tasa-chaveli_quest_2025, rey-montejo_discovery_2024}, sulfur bearing molecules \citep{2021A&A...646L...3C,cernicharo_sulphur_2021, 2024A&A...682L...4C, 2024A&A...688L..13C, remijan_missing_2025, sanz-novo_abiotic_2025, araki_detection_2026}, high-energy isomers \citep{cernicharo_discovery_2022, molpeceres_formic_2025, san_andres_first_2024, 2025A&A...703A..73Z, sanz-novo_conformational_2025} and cyclic species and PAHs \citep{cernicharo_italinfrared_2001, mcguire_detection_2018, Cernicharo2021_indene,mcguire_detection_2021, Cernicharo2021e, cernicharo_discovery_2022_ful,Cernicharo2021_benzyne, sita_discovery_2022, loru_detection_2023, cernicharo_discovery_2024, wenzel_detection_2024, wenzel_detections_2024, fuentetaja_discovery_2026, cabezas_discovery_2025, wenzel_discovery_2025}. The list above presents a non-exhaustive, yet representative sample of recent detections highlighting the diversity of chemical families.

Interestingly, a wide variety of these detections have been made in specific astronomical sources, of very different nature. For instance circumstellar envelopes (e.g. IRC+10216), star formation regions (Sgr-B2(N), NGC 6334I, IRAS 16293–2422), and molecular clouds (e.g. G+0.693-0.027, TMC-1). The latter case, and in particular the TMC-1 dark cloud, has been the most prolific source in terms of detections, with more than 120 detections (including species known in other sources) since 2018, an increase of approximatedly a 250\% in a short timespan. 

The TMC-1 dark cloud is a cold, dense region of the ISM located in the Taurus molecular cloud complex, that harbors the richest molecular inventory in a single source to date. As a cold, dense region TMC-1 is known to be one of the test beds for astrochemical models \citep{agundez_tmc-1_2026} and the prototypical source for the detection of new molecules, with detections in this source serving as stepping stones to the search in other sources \citep{burkhardt_ubiquitous_2021, 2023A&A...677L..13A}. The impressive rate of detections in TMC-1 has been driven by recent surveys, most notably the \textsc{Quijote} \citep{cernicharo_quijote_2022}, and \textsc{Gotham} \citep{McGuire2020} line surveys.\footnote{\textsc{Quijote}: Q-Band Ultrasensitive Inspection Journey to the Obscure TMC-1 Environment; \textsc{Gotham}: GBT Observations of TMC-1: Hunting Aromatic Molecules} 

\begin{figure*}
    \centering
    \includegraphics[width=\linewidth]{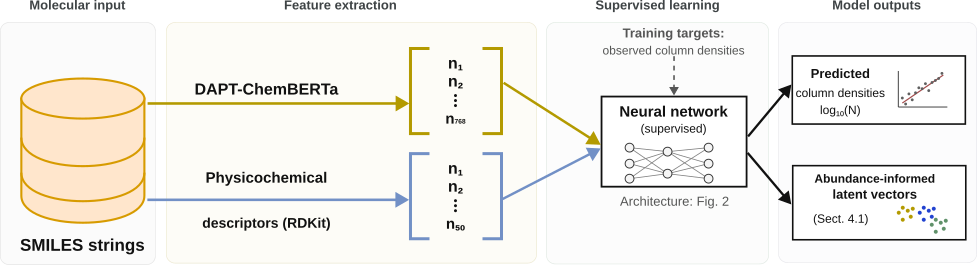}
    \caption{\rev{Schematic representation of the general strategy followed in this work. The specific architecture of the supervised learning model is later summarized in Figure \ref{fig:nn}.}}
    \label{fig:scheme}
\end{figure*}

\rev{The rapid expansion of molecular inventories has highlighted important challenges in our understanding of interstellar chemistry. While the continuous detection of new molecules stimulates experimental and theoretical efforts to understand their formation and destruction pathways, these studies struggle to keep pace with the number of detections. Consequently, many recently discovered species still lack well-established formation pathways or have not yet been incorporated into chemical networks. At the same time, identifying new molecular candidates is becoming increasingly challenging, as their selection relies heavily on chemical intuition and the availability of laboratory spectroscopic data. This is particularly relevant for unstable or highly reactive species, whose laboratory characterization can be difficult, summed to the high spectroscopic accuracy required for astronomical identification \citep[e.g.,][]{silva_discovery_2023, fuentetaja_space_2024}. Astrochemical models provide the fundamental framework for interpreting molecular observations, understanding the chemistry of the ISM, and suggest new candidates for detection. Yet, their predictive capabilities remain constrained by incomplete chemical networks, uncertainties in kinetic parameters, and observational limitations. Consequently, our picture of interstellar chemistry remains strongly influenced by the molecules we can detect, characterize, and model.
} 

\citet{lee_machine_2021}, on whose work the present manuscript builds, identified several gaps in our current understanding and prediction of astrochemical detections and proposed a data-driven framework inspired by analogous approaches in other scientific disciplines. Since its publication, this methodology has been applied in several studies \citep{scolati_explaining_2023, fried_implementation_2023, remijan_astronomical_2023, toru_shay_exploring_2025, toru_shay_bridging_2025,li_chemical_2026}. The general strategy consists of compiling a molecular inventory for a given source, encoding molecular structures into numerical representations using unsupervised machine-learning techniques such as \texttt{mol2vec} \citep{jaeger_mol2vec_2018}, \texttt{vicgae} \citep{lee_language_2021, scolati_explaining_2023}, or Morgan fingerprints \citep{rogers_extended-connectivity_2010}, and subsequently training a supervised model to relate these representations to observed molecular abundances. The resulting model can then be used to explore large chemical databases, such as \textsc{Zinc} \citep{irwin_zinc20free_2020}, to identify promising candidates for future detections or for astrochemical modelling. Despite its success, including its contribution to the detection of new interstellar molecules \citep[like \ce{CH3OCH2CH2OH}]{fried_rotational_2024}, candidate selection is typically based on proximity in the original embedding space and therefore primarily reflects structural molecular similarity rather than astrochemical relevance. Although chemical structure may, and in many cases certainly does, correlate with astrochemical relevance, this relationship does not hold uniformly across chemical space, as demonstrated later in this work. 

Here, we introduce a new framework that retains the general philosophy of \citet{lee_machine_2021} while introducing several methodological advances. First, rather than focusing exclusively on predicting molecular column densities, we use the latent representations learned during molecular-abundance prediction as the primary output of the model. Unlike scalar abundance predictions, these embeddings can be used for further processing, e.g. clustering or classification, enabling molecules to be grouped according to abundance-related patterns learned from the data. Second, we incorporate transformer-based chemical language models, which have proven highly effective at learning chemically meaningful representations directly from molecular strings \citep{chithrananda_chemberta_2020}, and adapt them to the astrochemical domain through domain-specific pretraining, resulting in improved predictive performance on both in distribution and out of distribution molecular inventories. Finally, we employ a multibranch neural-network architecture that jointly learns from molecular embeddings and physicochemical descriptors, allowing complementary structural and molecular information to be integrated within a single supervised representation. \rev{The flexibility of neural networks} for integrating multiple streams of information provides a natural framework for future extensions of the model, including the incorporation of three-dimensional molecular information, such as conformational isomerism. Over the course of the submission of this manuscript, \citep{fried_automated_2026} reported a complementary framework in which molecular embeddings are incorporated into an automated spectral-line assignment procedure to evaluate the chemical relevance of candidate carriers. Their approach further exploits the regions of chemical space occupied by the inferred molecular inventory to generate and rank potential targets for subsequent investigation. This development illustrates the increasing role of learned molecular representations not only in abundance prediction, but also as quantitative proxies for astrochemical context in observational analyses.

Our ultimate goal in this work is to present a new strategy for identifying correlations in chemical space while placing astrochemical considerations at the forefront. To this end, a substantial portion of the manuscript is devoted to describing the methodology and model training, presented in Sections \ref{sec:methodology} and \ref{sec:model_training}. In Section \ref{sec:applications}, we explore two \rev{preliminary} applications of the model. First, we investigate molecules that are deemed as astrochemically related as a preliminary step towards the identification of promising candidates for future detections. Second, we examine how recently detected molecules in TMC-1 have influenced our understanding of the source's chemical inventory. Finally, we conclude by discussing the main limitations of the model and summarizing the principal findings of this study.

%--------------------------------------------------------------------
\section{Methodology} \label{sec:methodology}

As it was mentioned in the introduction, our methodology builds upon the general strategy introduced by \citet{lee_machine_2021}. \rev{The general strategy of their work consists of combining unsupervised numerical representations of molecular structures with supervised machine learning to establish relationships between molecular structure and observed abundances, ultimately identifying promising candidates for interstellar detection.} 

A general scheme of our approach is shown in Figure \ref{fig:scheme}. A list of scientific libraries and acronyms used in this work is provided in Appendix \ref{sec:acronyms}.

\subsection{Preprocessing and molecule codification}

\subsubsection{General strategy to vectorize astrochemical molecules} \label{sec:featurize} 

The first step in our machine-learning pipeline consists of encoding molecular structures into numerical representations suitable for machine learning. To this end, we represent each molecule using the \textsc{SMILES} (Simplified Molecular Input Line Entry System; \citealt{weininger_smiles_1988}) notation, a text-based language that encodes molecular structures as character strings. For example, methane is represented as ``C'', carbon dioxide as ``O=C=O'', and acetonitrile as ``CC\#N''. These \textsc{SMILES} strings are then tokenized into sequences of chemical symbols, such as ``C'' for a carbon atom, ``='' for a double bond, and ``\#'' for a triple bond. We construct molecular embeddings by applying natural language processing (NLP) techniques to these tokenized sequences through domain-adaptive pretraining (DAPT; \citealt{gururangan_dont_2020}) of the \texttt{ChemBERTa} transformer model \citep{chithrananda_chemberta_2020}, an architecture designed to learn chemical information directly from \textsc{SMILES} representations. The model is based on the \texttt{RoBERTa} architecture \citep{liu_roberta_2019}, an optimized transformer language model originally developed for natural language understanding. Transformer training uses a masked language modeling (MLM) objective, in which a fraction of the tokens in each molecular string is masked, and the model is trained to reconstruct them from the surrounding context. This \rev{unsupervised} training allows the model to capture statistical and structural relationships across large molecular datasets and to encode molecules into vectors, referred to as molecular embeddings. These embeddings are obtained by averaging the activations of the final hidden layer of the \texttt{ChemBERTa} network over all tokens representing a molecule, resulting in a fixed-length vector that captures the molecular features learned by the model. These embeddings provide a compact representation of molecular structure and composition and constitute the initial feature space for the supervised learning tasks performed in this work. However, the original \texttt{ChemBERTa} models are predominantly trained on drug-like molecules and therefore reflect the chemical distribution of terrestrial organic chemistry, a limitation already acknowledged by previous studies \citep{lee_machine_2021}, with different protocols. \rev{To adapt the embedding space to the astrochemical regime, we performed DAPT using a dataset of molecules relevant to interstellar chemistry. Since existing molecular databases do not adequately represent the diversity of interstellar species, we generated a dedicated molecular inventory to provide suitable training data for this adaptation. The protocol to generate this large tailored astrochemistry-like inventory of molecules, named \texttt{Chaos}, is described in Appendix \ref{sec:app:chaos}.} As a starting point, we adopted a publicly available \texttt{ChemBERTa} model pretrained on approximately 100 million SMILES extracted from the \textsc{Zinc-20} database \citep{irwin_zinc20free_2020}.\footnote{\url{https://huggingface.co/DeepChem/ChemBERTa-100M-MLM}} Starting from this converged model, we continued the masked-language training for a small number of epochs (two in the present work).  This procedure allows the embedding space to be reoriented toward the chemical domain of interest, astrochemistry, while preserving the general chemical knowledge acquired during large-scale pretraining, as demonstrated in Appendix \ref{app:a}. After fine-tuning, the value of the loss function used to assess model performance remains comparable to that obtained prior to fine-tuning, indicating that domain adaptation does not compromise the overall predictive capability of the model. Following training, each molecule is encoded as a fixed-length vector representation of dimension \rev{768}. All the references to these embeddings in the supervised learning procedure showcased later in this work are referred with an ``emb'' subscript. 

\subsubsection{Addition of physicochemical descriptors}

As indicated in Section \ref{sec:featurize}, \rev{a constant dimension feature vector usable in the supervised learning step, requires a mean-pooling (average of the activations at the token-level)}. Through this procedure, some molecular information is lost, mainly in terms of molecular size, weight, or number of heteroatoms. To palliate this issue, and thanks to the flexibility allowed by using a neural network for supervised learning, we incorporate physicochemical descriptors as an additional input to the model. We include 50 physicochemical descriptors, obtained using the \texttt{RDKit} library \citep{rdkit}. The list includes topological (e.g. Morgan fingerprint, Labute), electronic (e.g. number of radical electrons, number of valence electrons), \rev{structural and compositional} (e.g. molecular weight, fraction of sp$^{3}$ carbon) and purely chemical (e.g. number of aromatic rings, number of OH groups) descriptors. A full list of the descriptors used in this work is available in the repository accompanying this work. The choice of the descriptors was made based on their relevance in the context of interstellar chemistry, from the list available in \texttt{RDKit}, and was not optimized. Nevertheless, its relevance in the subsequent supervised learning procedure is ultimately determined by the model itself, as the importance of these descriptors is learned and included in a parallel input layer to the \rev{embedding one. Both layers are later concatenated.} All the parameters referred to this input layer are denoted with the ``phys'' subscript in the supervised learning procedure. 

\subsection{Model architecture}

\begin{figure}
    \centering
    \includegraphics[width=\linewidth]{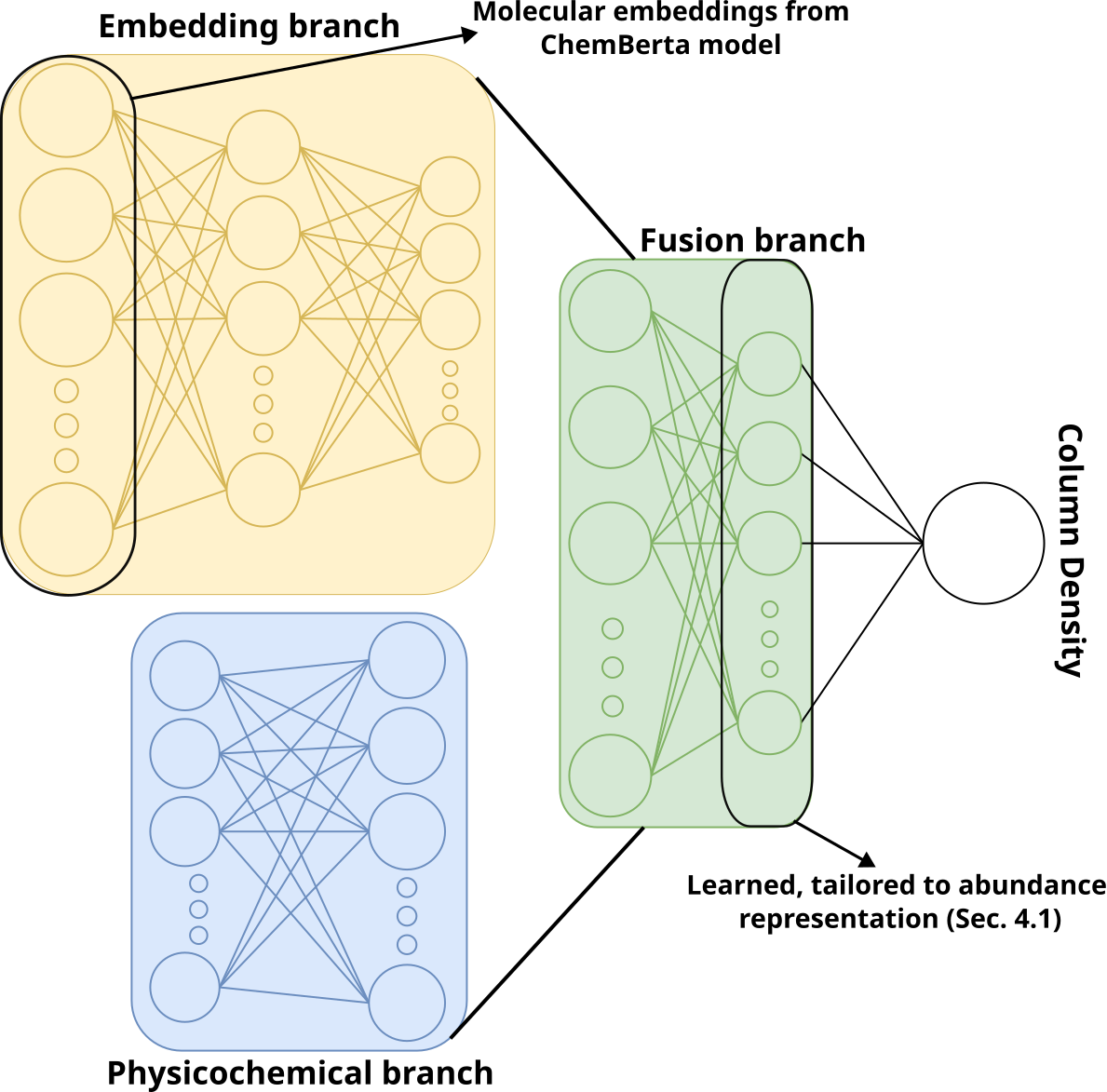}
    \caption{Schematic representation of the architecture of the neural network model used for the supervised learning procedure.}
    \label{fig:nn}
\end{figure}

The final supervised learning model arises from the combination of the two types of features described in the previous sections, the finetuned \texttt{ChemBERTa} molecular embeddings and the physicochemical descriptors. The regressor that we use to predict the target function, which in our case is the logarithm of the column density, is a fully connected multilayer perceptron neural network. The optimized structure of the neural network architecture is described in Section \ref{sec:architecture}. On a more general level, the model consists of two parallel branches that are later concatenated in a new layer, named ``fusion'' (subscript ``fus''), to produce a combined representation of the molecule. A given set of molecules, specifically the TMC-1 molecular inventory considered in this work, is randomly divided into training (80\%) and validation (20\%) sets. The training set is used to fit the model, whereas the validation set is used to evaluate its performance on unseen data. For each task, we train several models using different random train–validation splits and report the average of their predictions. The number of models used for each task is specified in the corresponding section. All input vectors are standardized before passing through the network to have zero mean and unit standard deviation in the training set, fitted independently for each branch. The neural-network model is constructed using \texttt{Keras} \citep{chollet2015keras} with \texttt{Tensorflow} as backend \citep{tensorflow2015-whitepaper} while training/validation split and standardization of the data is handled by \texttt{scikit-learn} \citep{scikit-learn}.

Before concatenation, the embedding branch passes through two fully connected layers, while the physicochemical descriptor branch passes through a single fully connected layer. The fusion layer, is then passed through an additional fully connected layer before the final output. A scheme of the general architecture is shown in Figure \ref{fig:nn}.

Because we are working with a small training set, several regularization techniques are applied to the model, with their specific hyperparameters optimized as described in Section \ref{sec:architecture}. Briefly, the regularization techniques that we use are neuron dropout, layer normalization, L2 regularization and Gaussian noise addition. These techniques are used through the equivalent \texttt{Keras} implementations. Neuron dropout consists in randomly deactivating a fraction of the neurons in a given layer during training, which prevents the model from relying too heavily on specific neurons and encourages it to learn more robust features. Layer normalization is a technique that re-normalizes the activation functions of a given layer across the features, which can help to stabilize and accelerate training, specially in layers where a significant change in the distribution of the input features is expected. L2 regularization consists in adding a penalty term to the loss function that is proportional to the square of the weights of the model, which encourages the model to learn smaller weights. This is useful to avoid that some transformation within the neural-network architecture dominate training, resulting in overfitting. Finally, Gaussian noise addition consists in adding random noise drawn from a Gaussian distribution to the input features during training, which can help to improve the robustness of the model. It is important to indicate that not only the specific hyperparameters connected to these regularization techniques are optimized (e.g. 0.4 dropout probability in a given layer), but also whether a given regularization technique is applied or not in a given layer. The model is trained minimizing a Huber loss function \citep{huber_robust_1964}. The transition parameter ($\delta_{\mathrm{Huber}}$) is also optimized as a hyperparameter. Because we later perform hyperparameter optimization with mean squared error as target (MSE; Section \ref{sec:architecture}), we tend to favor a loss function that behaves more similarly to MSE than to the mean absolute error (MAE). \rev{Finally, to mitigate overfitting, the training procedure also incorporates the \texttt{keras} EarlyStopping and ReduceLROnPlateau procedures, which monitor the validation loss to stop training when further improvements are no longer observed and adaptively reduce the learning rate when convergence slows. The parameters governing these procedures were kept fixed rather than included in the hyperparameter optimization.}

As we will discuss further in the text, our neural network model provides more than predictions of the column density, which is the target variable. It also encodes meaningful (in terms of molecular column densities) representations of the input features. These representations capture the underlying structure of the data and can be exploited. We use them to visualize the molecular inventory (Section \ref{sec:inference}) and project new molecules into that space, which is a central objective of this work, rather than performing direct column density regression. More broadly, the information contained in the hidden layers could also be used for model interpretability and explainability analyses, although this is postponed for future work.

\subsection{Model optimization} \label{sec:architecture}

\begin{table}
\caption{Hyperparameter configuration of the neural network model.}
\label{tab:model_hyperparameters}
\centering
\begin{tabular}{lcc}
\hline\hline
Hyperparameter & Starting value & Optimized value \\
\hline

$N_{\mathrm{emb},1}$ & 256 & 512 \\
$N_{\mathrm{emb},2}$ & 128 & 256 \\
$N_{\mathrm{phys}}$ & 32 & 64\\
$N_{\mathrm{fus}}$ & 128 & 198\\

$\sigma_{\mathrm{emb}}$ & 0.05 & 0.11\\
$\sigma_{\mathrm{phys}}$ & 0.05 & 0.09\\

$p_{\mathrm{emb},1}$ & 0.15 & 0.30\\
$p_{\mathrm{emb},2}$ & 0.15 & 0.30\\
$p_{\mathrm{phys}}$ & 0.15 & 0.25\\
$p_{\mathrm{fus}}$ & 0.15 & 0.10 \\

$\mathrm{LN}_{\mathrm{emb},1}$ & False & False\\
$\mathrm{LN}_{\mathrm{emb},2}$ & False & False \\
$\mathrm{LN}_{\mathrm{phys}}$ & False & True \\
$\mathrm{LN}_{\mathrm{fus}}$ & False & False\\

$\lambda_{\mathrm{emb},1}$ & 1$\times 10^{-4}$ & 8.4$\times$10$^{-3}$\\
$\lambda_{\mathrm{emb},2}$ & 1$\times 10^{-4}$ & 4.6$\times$10$^{-5}$\\
$\lambda_{\mathrm{phys}}$ & 1$\times 10^{-4}$ & 4.6$\times$10$^{-4}$\\
$\lambda_{\mathrm{fus}}$ & 3$\times 10^{-4}$ & 4.0$\times$10$^{-3}$\\

$\phi$ & Swish function & GELU function \\
Optimizer & Adam & NAdam\\
$\eta$ & 3$\times 10^{-4}$ & 7.4$\times$10$^{-4}$ \\
$\delta_{\mathrm{Huber}}$ & 0.5 & 2.0 \\

\hline
\end{tabular}
\tablefoot{$N$ denotes the number of neurons. Subscripts ``emb'', ``phys'', and ``fus'' refer to parameters associated with the embedding input, the \rev{physicochemical} descriptors, and their combined representation, respectively. $\sigma_{\mathrm{emb}}$ is the standard deviation of the Gaussian noise applied to the inputs. $p$ denotes dropout probabilities. $\mathrm{LN}$ indicates whether layer normalization is applied. $\lambda$ denotes L2 regularization strengths. $\phi$ is the activation function, $\eta$ the learning rate, and $\delta_{\mathrm{Huber}}$ the transition parameter of the Huber loss function. Optimizer is the optimization algorithm used for training, selected between Adam, RMSProp, and Nadam.}
\end{table}

\rev{The relatively shallow architecture was motivated by our experience with neural-network potentials, where increased depth can improve fitting at the expense of generalization. Although the effect of network depth was not systematically explored here, the number of neurons per layer was optimized to balance model complexity and predictive performance.} Even though relatively small, the model, for an arbitrary architecture, has many hyperparameters influencing its performance. The amount of trainable parameters is determined mostly by the size of the hidden layers. Besides, as already discussed above due to the scarcity and heterogeneity of the dataset, several regularization techniques are applied. In summary, the goodness of the model fit relies on a number of hyperparameters significantly higher than other reported models. Therefore, hyperparameter optimization is mandatory in order to obtain the best model for our chosen architecture. We perform hyperparameter optimization using the Bayesian optimization routines implemented in \texttt{Optuna} \citep{optuna_2019}.  The optimization is stopped when the last 100 iterations do not improve upon the best-performing iteration, using the initial set of hyperparameters listed in Table \ref{tab:model_hyperparameters}. The hyperparameter optimization is carried out using ten different validation splits, obtained with Monte Carlo cross validation. For each optimization cycle ten models with a different random split between training and validation data are used, keeping a 0.80/0.20 partition. The optimization is carried out minimizing the following score function:
\begin{equation}
   \rm{Score} = \overline{\rm{MSE}}_{\mathrm{val}} + 0.25\sigma_{\mathrm{MSE}_{\mathrm{val}}} + 0.25\overline{\rm{MSE}}_{\mathrm{train}},
\end{equation}
across the different validation splits. The hyperparameter optimization is carried out with the full dataset of TMC-1 at the date of January 2026 containing 184 molecules.\footnote{Strictly, the number of molecules detected in TMC-1 is slightly higher, but our model cannot yet handle molecular conformers, like cis-HCOOH \citep{molpeceres_formic_2025} or gauche/anti-\ce{CH2CHCH2CN} \citep{cernicharo_discovery_2022}, to give some of the examples.} The SQLite database for the full optimization, as well as the list of SMILES and column densities of TMC-1 are available in the Zenodo repository accompanying this article \url{https://zenodo.org/records/23531603}.

\section{Model training} \label{sec:model_training}
The final set of hyperparameters is gathered in Table \ref{tab:model_hyperparameters}, last column. The \texttt{Optuna} library also provides a ranking of the importance of the different hyperparameters. In our case, the most important hyperparameters are the L2 regularization strengths ($\lambda$), the dropout probabilities ($p$), the Gaussian noise standard deviations ($\sigma$) and the learning rate ($\eta$). \rev{For example, the importance of the regularization hyperparameters is consistent with the limited size and heterogeneity of our dataset, as regularization aids the model from becoming too biasedtowards molecular features or species. This also suggests that increasing the amount and diversity of training data will improve model robustness and predictive performance, since the current model, and others in the literature, operate in a clear data scarcity regime.}

\begin{figure}
    \centering
    \includegraphics[width=0.85\linewidth]{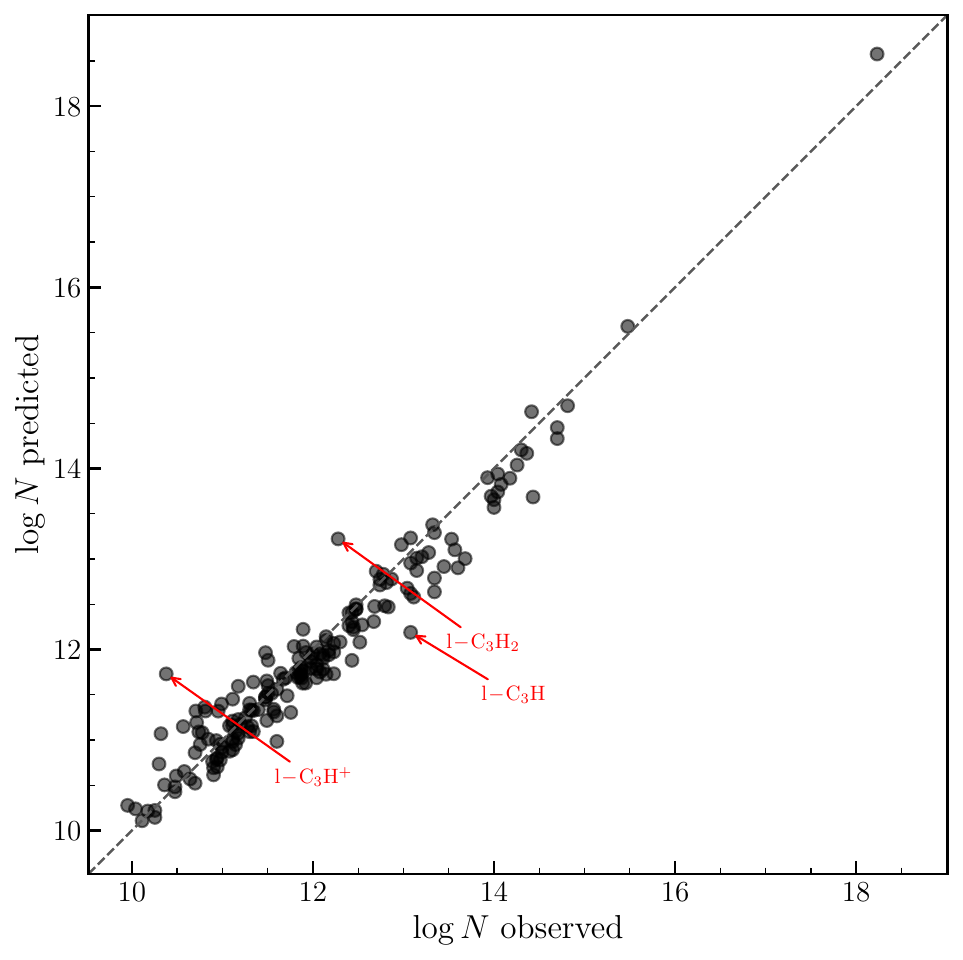}
    \caption{Parity plot for the production model. In red the three worst behaving points, namely \ce{l-C3H+}, \ce{l-C3H} and \ce{l-C3H2}. The results are obtained as the average of three models using different initialization seeds, including different train and validation splits.}
    \label{fig:correlation}
\end{figure}

\begin{figure}
    \centering
    \includegraphics[width=\linewidth]{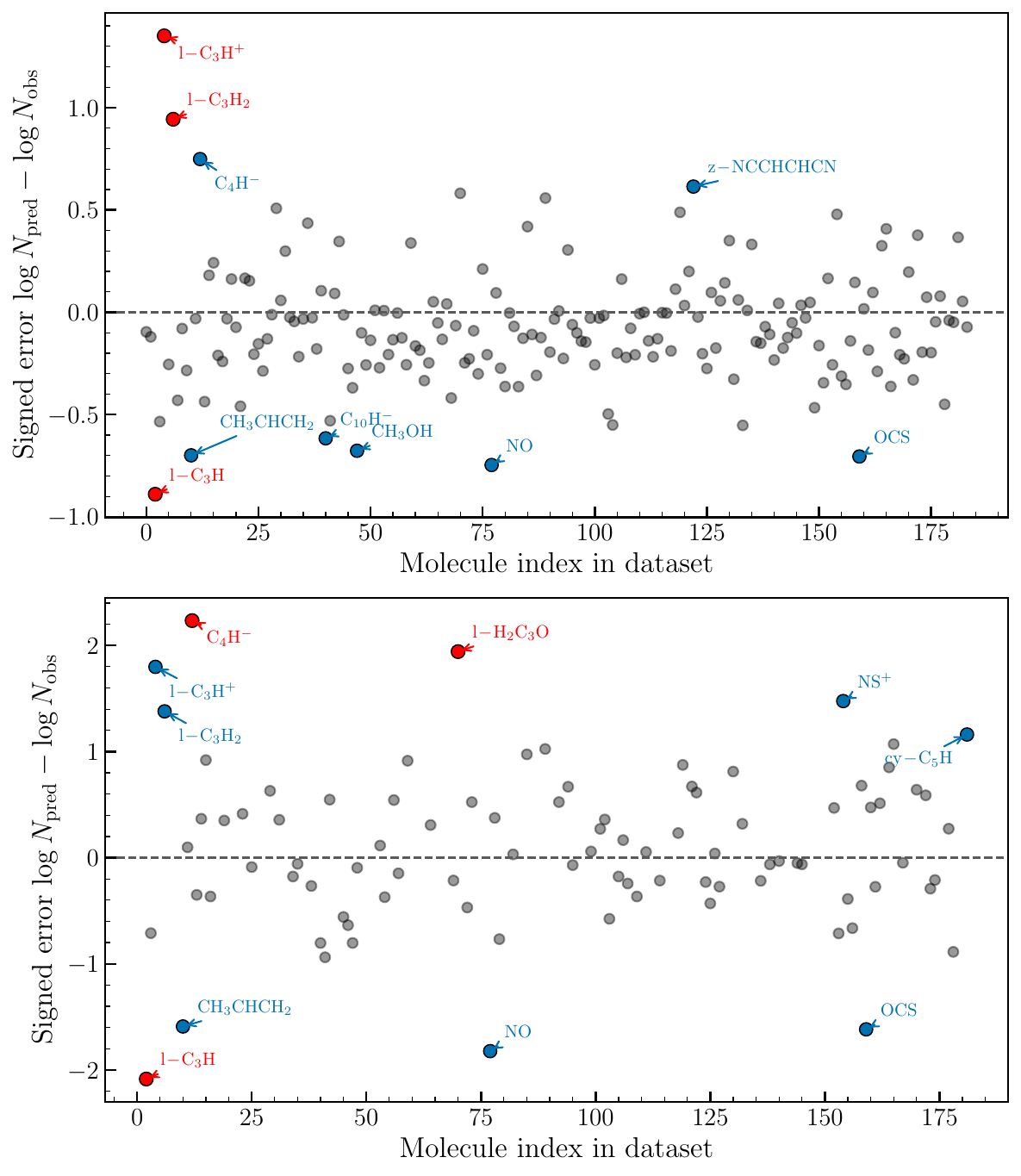}
    \caption{Signed absolute deviation between observed and predicted column densities. Top. Train + validation subset averaged runs from different models, consistent with Figure \ref{fig:correlation}. Bottom. Only molecules in the validation subset, averaged by the amount of times they appear in such subset. Red points indicate the three worst behaving molecules whereas blue points indicate the remaining worst behaving points up to ten.}
    \label{fig:outliers}
\end{figure}

\rev{The results obtained using three independent instances of the same model, with identical optimized hyperparameters (Table \ref{tab:model_hyperparameters}), but different random initializations and training–validation splits, are shown in Figure \ref{fig:correlation}. For each instance, the molecular inventory is randomly divided into training (80\%) and validation (20\%) subsets, and the reported metrics are averaged across the three instances.} For the combined training and evaluation set, the averaged metrics yields an MSE of 0.17$\pm$0.03, an MAE of 0.28$\pm$0.01, and an $R^{2}$ (coefficient of determination) value of 0.89$\pm$0.02. When evaluated exclusively on the validation sets, the corresponding values are 0.58$\pm$0.17, 0.59$\pm$0.08, and 0.51$\pm$0.17, respectively. The reported uncertainties come from the standard deviation of the three trained models. Both the MSE and MAE are expressed in $\log(N)$. \rev{In particular, the validation MAE of 0.59 corresponds roughly to an uncertainty in the predictions for unseen molecules of 10$^{0.59}$, that is a factor four between predicted and observed column densities. This accuracy is primarily representative of molecules within the chemical domain sampled by the training inventory, and should not be assumed to hold for out-of-distribution molecules (further discussed in Section \ref{sec:applications}).} The MSE and $R^{2}$ values can be compared with those reported by \citet{toru_shay_exploring_2025}, who obtained an MSE of 0.19 and an $r^{2}$ value of 0.81. However, this comparison should be interpreted with caution because the two studies differ in several methodological aspects, including for example the cross-validation strategies employed, bootstrap resampling, the use of an an average of models in our reported data or simply because their metrics are reported with respect a less complete dataset to compare \rev{with the original model of 2021 \citep{lee_machine_2021}}. In Section \ref{sec:explainability} we also show how our predictions worsen when using a less complete dataset.

\rev{Diatomic species, particularly carbon monoxide (CO; $N = 1.7 \times 10^{18}$ cm$^{-2}$; \citealt{Pratap1997}), are represented among the most abundant molecules in the dataset, with CO appearing at the upper end of the diagonal in Figure \ref{fig:correlation}. CO exhibits an atypical bonding pattern that is poorly represented in the molecular inventory used for training. For instance, its SMILES representation, \texttt{[C-]\#[O+]}, is unique within the dataset. These characteristics introduce two potential difficulties: first, its extreme column density significantly expands the regression domain beyond the typical range ($10^{10}$–$10^{14}$ cm$^{-2}$), second, its unusual molecular representation, combined with an extreme target value, may have a significant impact the training procedure and worsen the predictive performance for other molecules. This does not necessarily imply that its individual column density is poorly reproduced when included in the training set. A comparable situation arises for \ce{C24H11CN}, a recently detected species \citep{wenzel_discovery_2025}, which contains seven fused benzene rings, is currently the largest molecule detected in TMC-1, and represents an extreme case in terms of molecular size within the dataset. For these reasons, \ce{CO} and \ce{C24H11CN} were systematically retained in the training subset and excluded from validation. From our tests, it is not clear if removing them from the training subset also improves generalization, its evaluation requiring an in-depth investigation with out of distribution molecules. Nevertheless, we decided to include them in the training subset as a way to constrain the range of molecular abundances and structures represented during training, which is later suggested by our out of distribution predictions (Section \ref{sec:inference}).}

Both in Figure \ref{fig:correlation} and Figure \ref{fig:outliers}, where we show the outliers both in train/validation and solely in validation, we have a common set of worst-behaving points, for example l-\ce{C3H+}, l-\ce{C3H}, l-\ce{C3H2}, l-\ce{H2C3O}, OCS, NO, \ce{C10H-} or \ce{C4H-}. While we certainly cannot understand the reason why every specific molecule is predicted worse than the others, we can rationalize the behavior for several of them. Starting with l-\ce{C3H+}, our model consistently overpredicts its abundance by two orders of magnitude in validation and by about one and a half orders of magnitude in training plus validation. Interestingly, l-\ce{C3H+} is a cation of the \ce{C_xH_y+} family that reacts with \ce{H2} through the reactions \citep{1993ApJ...417..181M}:
\begin{align}
   \ce{l-C3H+ + H2 &-> l-C3H2+ + H} \\
   \ce{l-C3H+ + H2 &-> l-C3H3+ + h\nu} \\
   \ce{l-C3H+ + H2 &-> c-C3H2+ + H} \\
   \ce{l-C3H+ + H2 &-> c-C3H3+ + h\nu}
\end{align}
Reactions with \ce{H2} rarely dominate the destruction of molecules in dark clouds since they normally involve activation barriers. The exceptional reactivity of \ce{l-C3H+} is therefore a rarity that cannot be learned by the model, there is simply not enough information available to capture such an exception. Moving forward to the l-\ce{C3H}/l-\ce{C3H2} pair, the molecules are either over or under predicted. We believe the reason for the specific deviations of this pair lies in a selective electron recombination reaction affecting the precursor, c-\ce{C3H3+} \citep{2017MNRAS.470.4075L}, favoring \ce{H2} elimination over H, producing favorably \ce{l-C3H} (underpredicted) over \ce{l-C3H2} (overpredicted). The  explanation for next molecule, l-\ce{H2C3O}, recently detected in TMC-1 \citep{esplugues_discovery_2026}, is much clearer. Our model overpredicts l-\ce{H2C3O}. This molecule is also particular from a chemical point of view and has been the subject of extensive debate in the literature \citep{loomis_investigating_2015, loison_interstellar_2016, shingledecker_case_2019,esplugues_discovery_2026} because it represents one of the clearest cases of the breakdown of the minimum energy principle (MEP) for interstellar isomers \citep{lattelais_interstellar_2009}. Briefly, \ce{l-H2C3O} is the most stable isomer of the molecular formula, yet it has eluded detection until 2026 and in low abundances, while the less stable isomers were detected much earlier. Several explanations have been given for that anomaly, including fast hydrogenation on grains or selective dissociative recombination. In any case, our neural network model reaches the same intuitive conclusion as the chemical literature, in an unbiased manner, hence leading to a clear overprediction for purely chemical reasons. Finally, the overprediction of the small anion \ce{C4H-}, and the underprediction of the large anion \ce{C10H-} finds its origin in the well known increment in reactivity with increasing size \citep{herbst_dissociative_2011}. The abundance of small anions is lower because their formation rate is very small, and the opposite occurs in large anions. This is the reason why the model later infers small anions as good candidates for detection in TMC-1 (Section \ref{sec:inference}.) This effect that cannot be inferred from our data structure and that shows a nice chemical explanation for the shortcomings of our model. Some other are badly predicted simply because the model does not have sufficient information by construction, like OCS or NO, with unique, hard to interpolate, structures.

It would be a futile effort to rationalize the behavior of every single molecule, but the examples above show that the worst-behaving points are not random, rather they correspond to molecules with particular and hard-to-predict chemical reactivities. This is not only an indication that our model is learning meaningful chemical relationships, but also that its failures can be used to identify interesting chemical cases that deserve further attention, either from an observational or a theoretical point of view. 

\section{Applications} \label{sec:applications}

Having outlined the methodology and the training of the model, including its training and validation metrics, the two applications considered in this work. As in the seminal model introduced by Lee and its subsequent revisions \citep{lee_machine_2021, fried_implementation_2023, scolati_explaining_2023, toru_shay_exploring_2025,toru_shay_bridging_2025, li_chemical_2026, fried_automated_2026}, the most immediate application of the model is the exploration of molecules not previously seen during training. While we incorporate a similar exercise in this work, our focus is not only restricted to the identification of new, previously undetected species. Instead, we primarily seek to understand the characteristics that make a molecule astrochemically abundant. In particular, our protocol also highlights molecules that are expected to be abundant yet remain invisible to radioastronomical observations, such as \ce{H2}. The second application aims to quantify the paradigm shift produced by the recent detections in TMC-1 and their impact on our understanding of the chemistry of this source. To this end, we train the model using only the molecular inventory available prior to these detections and evaluate whether the learned representation is capable of rationalizing the presence of the newly detected species.

\subsection{Astrochemically favored out-of-distribution molecules} \label{sec:inference}

\subsubsection{Protocol to generate new astrochemical candidates}

\begin{figure*}[ht]
    \centering
    \includegraphics[width=\linewidth]{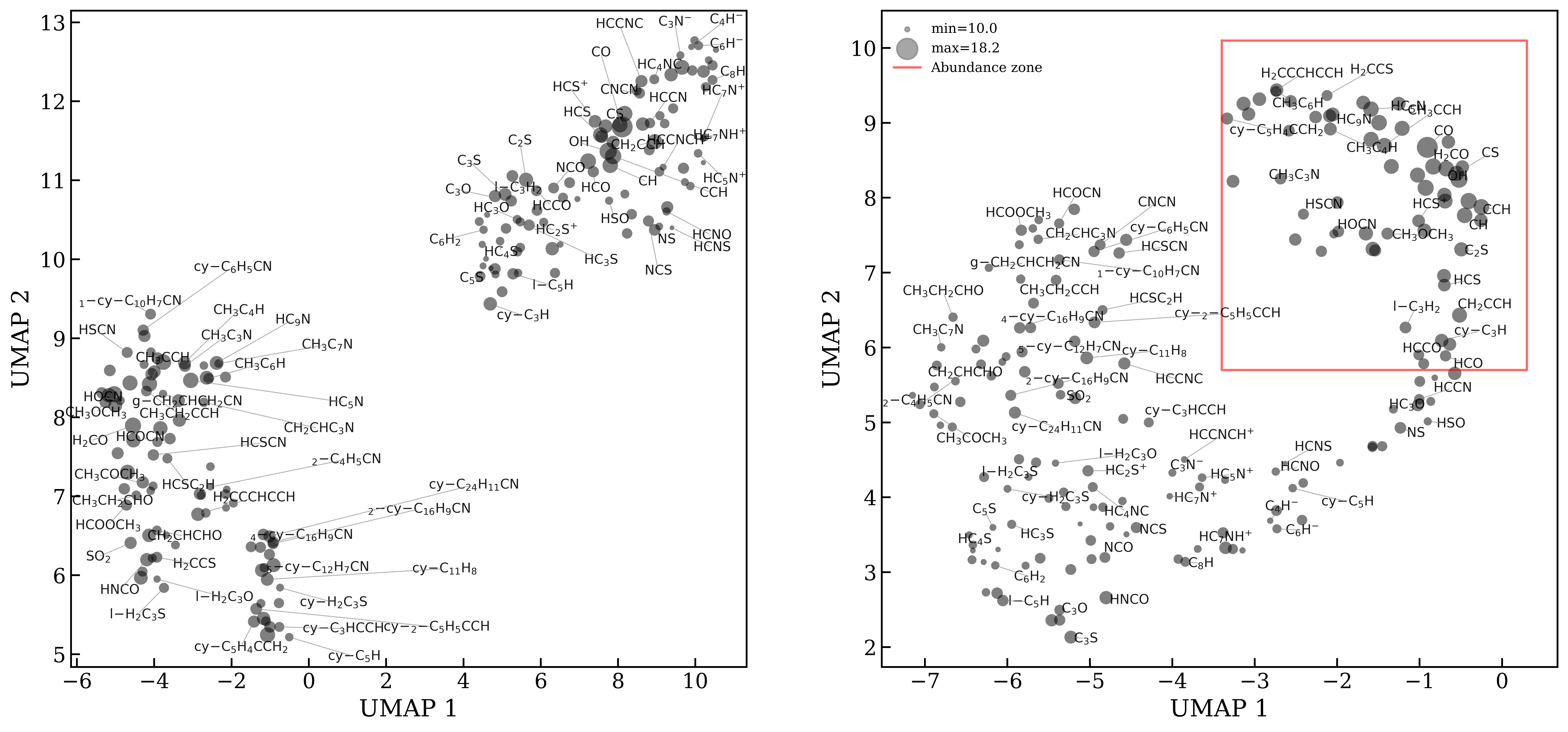}

    \caption{Left. Two-dimensional UMAP representation of the TMC-1 inventory using the original \texttt{ChemBERTa} input vectors. Right. Same representation but using the last hidden layer of the neural netwok model. The red rectangle highlights the region of the UMAP representation where highly abundant molecules are located.  The different dot sizes represent the observed column density of each molecule.
    }
    \label{fig:umap_representation}
\end{figure*}

\begin{figure*}
    \centering
    \includegraphics[width=0.85\linewidth]{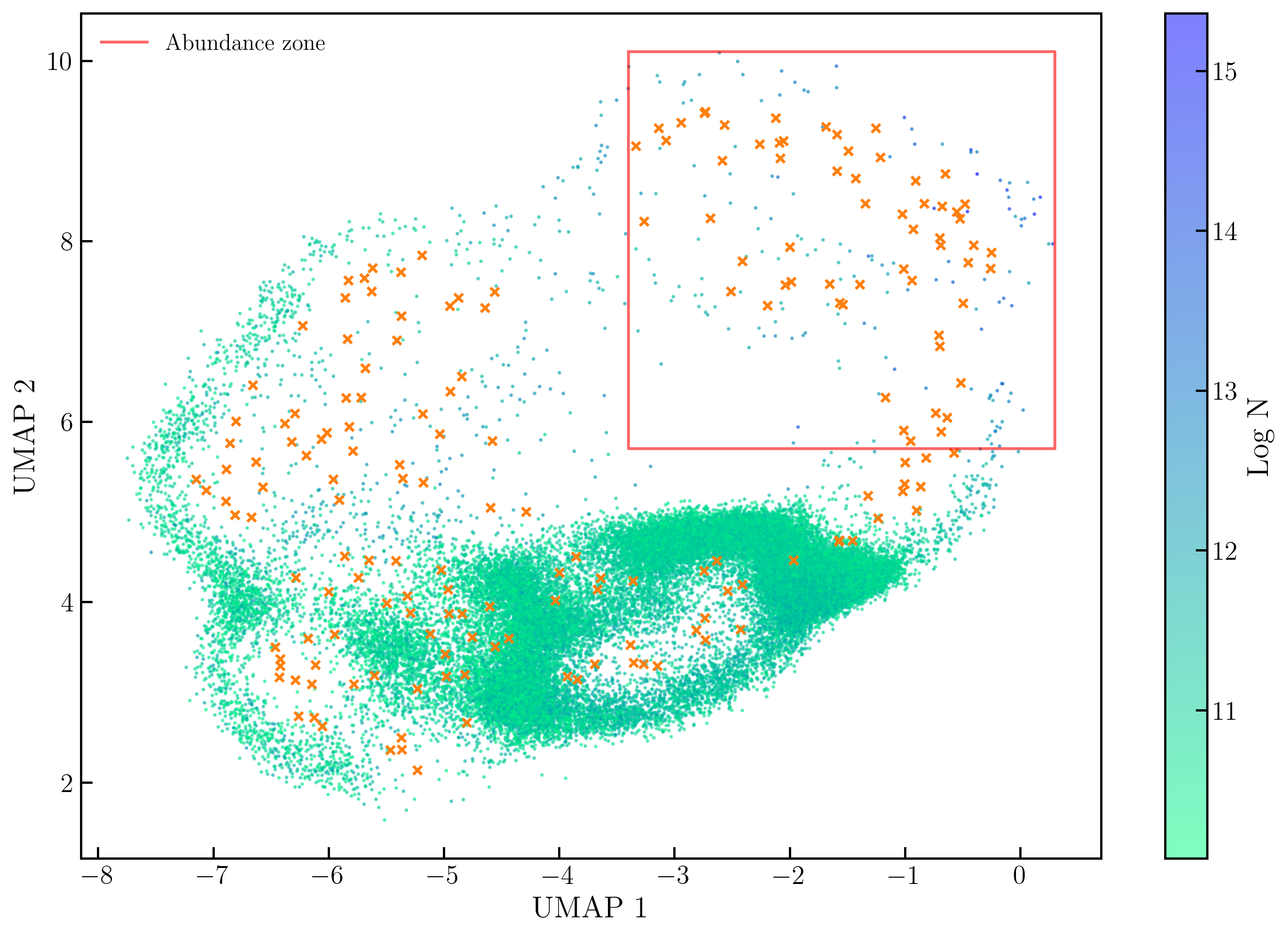}
    \caption{Projection of the model's latent vectors for the first 50,000 molecules generated with \texttt{Chaos} chaos protocol (\rev{Color coded by abundance and excluding CO to not saturate the color map}) onto the UMAP representation of the TMC-1 inventory (orange crosses). The red rectangle highlights the region of the UMAP representation where highly abundant molecules are located. The molecules within this region are considered promising candidates for future detections.}
    \label{fig:umap_clustering}
\end{figure*}

Our neural network model performs a regression that can be considered reliable only within the domain of the fitted function, namely the range of column densities spanned by the TMC-1 inventory (10$^{10}$–10$^{18}$ cm$^{-2}$). Therefore, for arbitrarily complex molecules, and in the absence of prior information establishing a meaningful lower bound on molecular abundances across chemical space,\footnote{In practice, the overwhelming majority of chemically possible molecules are expected to have negligible or effectively zero column densities.} the model must either interpolate within this sparse domain or extrapolate beyond it. Its predictions are therefore expected to become less reliable for complex, low-abundance species. \rev{This limitation gives rise to an apparent paradox}: the model is less reliable precisely for molecules that are absent or extremely rare in space, even though distinguishing such species from plausible interstellar molecules is one of the model's main purposes. We expect this limitation to be common to all models reported in the literature, although Bayesian approaches, such as those introduced by \citet{lee_machine_2021}, have the advantage of naturally providing uncertainty estimates, that might help in certain cases. We tested uncertainty estimation strategies such as ensemble averaging within our pipeline, but a more detailed investigation of methods to improve model performance below the detection threshold is deferred to future work. In the present study, rather than relying directly on uncertain low-abundance predictions, we identify promising detection candidates by evaluating their proximity, in the latent vectors trained on column densities, to already detected molecules, a strategy unique to our neural network architecture.

Our strategy for determining the proximity of potential candidates consists of two steps. In the first step, we represent the TMC-1 inventory using the task-specific latent vectors extracted from the last hidden layer of the supervised learning model, i.e. the activations of the last layer prior to column density regression. A single train/validation split is used in this task. Each latent vector is a 198-dimensional learned representation obtained by jointly processing the original \texttt{ChemBERTa} molecular embeddings and the physicochemical descriptors. Thus, these latent vectors are distinct from the input \texttt{ChemBERTa} embeddings,\footnote{The terminology ``latent vector'' is used to distinguish then from the input embeddings, although both are strictly learned embeddings.} as they encode a task-specific representation learned during supervised training. We first apply dimensionality reduction using the Uniform Manifold Approximation and Projection (UMAP) algorithm \citep{mcinnes_umap_2018}. Following \citet{lee_machine_2021}, we used UMAP to generate two-dimensional representations of the molecular inventory. In the first place, these representations were derived from two distinct feature spaces: the original molecular embedding space generated by \texttt{ChemBERTa}, and the latent vectors extracted from the last hidden layer of the supervised model, as described above. The results are shown in Figure \ref{fig:umap_representation}. In it, we represent the observed column density with different dot sizes. Interestingly, from the visualization of the inventory with the \texttt{ChemBERTa} embedding space as input, we can see a clear differentiation in chemical families, with molecules with similar chemical structure grouped together (e.g., linear cumulenes , O-bearing COMS, or cyclic species). However, while the chemical families are still sometimes grouped together in the representation of the inventory using the last hidden layer as input, the chemical structure is not the main driver of the representation, but rather the column density, as the model has learned a multidimensional vector representation strictly tied to it. This is evident by looking at the distribution of dot sizes in Figure \ref{fig:umap_representation} right panel, molecules grouped together by abundance. This structure is not apparent when the original \texttt{ChemBERTa} embedding space is used, for which the dot sizes are distributed more uniformly across the representation. In contrast, the task-specific latent space exhibits a clear separation according to molecular abundance. This is potentially the most impactful aspect of our work, as it opens up not only the possibility of overcoming the limitations in the regression of molecules, but also provides a basis to understand chemically related molecules and to rationalize the reasons for the presence of a given molecule in the ISM.

As shown in the right panel of Figure \ref{fig:umap_representation}, highly abundant molecules occupy a distinct region of the UMAP representation, highlighted by the red rectangle. We constructed latent vectors for the first 50,000 molecules generated with \texttt{Chaos} (Appendix \ref{sec:app:chaos}) and projected them using the same UMAP dimensionality-reduction procedure. These molecules were selected because they are the closest to the original TMC-1 inventory in the \texttt{Chaos} generation sequence. The resulting projection is shown in Figure \ref{fig:umap_clustering}. Most of the projected molecules lie outside the region occupied by highly abundant species, \rev{and with very small column densities}. We therefore hypothesize that model uncertainty increases with distance from this region. Although the uncertainty is expected to depend on the individual molecule, predictions for molecules located closer to the highly abundant region are likely to be more reliable. Accordingly, the following discussion considers only molecules that fall within the red rectangle in Figure \ref{fig:umap_clustering}. The purpose of this analysis is to illustrate a downstream application of the model rather than to provide a definitive list of candidate molecules. Although the resulting candidates are discussed below, they should be interpreted primarily as a demonstration of the model's potential. We also acknowledge that the manual selection of a region in the UMAP representation could be replaced in future work by a more robust clustering procedure, and that a higher-dimensional UMAP representation might provide a clearer separation between molecular populations. Nevertheless, we consider the protocol presented here to be best suited for illustrating our approach.

\subsubsection{Molecules selected by the model}

After filtering the molecules already present in the original dataset we end up with a list of 158 molecules. We present a selection of those in Table \ref{tab:smiles_candidate_molecules} for illustrative purposes, with the full list of molecular smiles provided as supplementary information in the accompanying repository.

\begin{table*}
\caption{Selected \rev{56} molecules from the predictions in \ref{sec:inference}. Molecules are not presented in order of abundance.}
\label{tab:smiles_candidate_molecules}
\centering
\resizebox{\textwidth}{!}{
\begin{tabular}{cccccccc}
\hline\hline

\ce{H2} & \ce{CH4} & \ce{C2H4} & \ce{C2H6} & \ce{H2O} & \ce{CO2} & \ce{C2H2} & \ce{H2O2} \\
\ce{cy-C6H6} & \ce{OH-} & \ce{CH3} & \ce{CH3+} & \ce{CH2NH} & \ce{HNO} & \ce{CH2(OH)2} & \ce{CH2} \\
\ce{cy-C10H8} & \ce{CH2CH} & \ce{C2NH}$^{*}$ & \ce{C2H-} & \ce{HCCCCH} & \ce{CH2CCH2} & \ce{CH3O} & \ce{N2H4} \\
\ce{NH2CN} & \ce{HCCNH2} & \ce{HCCOH}$^{*}$ & \ce{NCCN} & \ce{CH3NH2} & \ce{cy-C6H5OH} & \ce{CH3CCCH3} & \ce{CH2NH2} \\ 
\ce{CH2O} & \ce{CH2CN+} & \ce{CH3CHCHCCH} &  \ce{CH2C(CCH)CH3}$^{*}$ & \ce{C2-} & \ce{H2S2} & \ce{NH2CHS}$^{*}$ & 
\ce{CH2CHNH2} \\
\ce{CH2CHSH} & \ce{C4O} & \ce{HOCCOH}$^{*}$ & \ce{S3H2} & \ce{C2H5+} & \ce{CH3NOH} & \ce{CH2OH+} & \ce{NH2OH} \\
\ce{CHCCH} & \ce{l-C10H2} & \ce{CH2OH} & \ce{NH2CHNH} & \ce{CH2(NH2)(OH)} & \ce{CH3OCCH}$^{*}$ & \ce{HCCOCCH}$^{*}$ & \ce{CH3C(O)SH}$^{*}$ \\
\hline
\end{tabular}
}
\tablefoot{The table reports the raw prediction output, without excluding molecules lacking a permanent electric dipole moment. In addition, some predictions may correspond to species that have already been searched for unsuccessfully, reflecting both the incompleteness of the training dataset and the intrinsic bias towards detected molecules in the available observational data; see the main text for further discussion. We note that certain molecules cannot be generated through the substitution rules implemented in the \texttt{Chaos} protocol, like \ce{N2}, \ce{O2}, or other simple PAHs. A manual inspection of these molecules predict them to be equally abundant. Molecules marked with an asterisk ($^{*}$) are handpicked molecules for which dipole moments have been calculated, see text}
\end{table*}

Many of the molecules shown in Table \ref{tab:smiles_candidate_molecules} are well established molecules that are assumed to be amongst the most abundant interstellar molecules, but that lack permanent electric dipole moment, e.g., \ce{CH4}, \ce{C2H2}, \ce{C2H4}, \ce{CH3+}, \ce{CO2} or \ce{C6H6}. Interestingly, the model predicts \ce{H2} as one of our candidates without any prior knowledge that this molecule is known to be the most abundant one in the Universe. The model further predicts the unsubstituted polycyclic aromatic hydrocarbons (PAHs) benzene and naphtalene. In contrast, molecules such as \ce{N2}, \ce{O2}, or other simple PAHs do not appear among the candidates because they cannot be generated through the substitution rules implemented in the \texttt{Chaos} protocol. Altogether, these results indicate that the model is capable of extracting chemically meaningful information that was never explicitly encoded during training. The predicted set also contains several molecules that are known or expected to be abundant in the interstellar medium, but remain difficult to observe because of strong depletion onto dust grains, such as \ce{H2O} and \ce{HNO}. Likewise, the model exhibits a noticeable bias toward anions and carbene/carbyne-like species, which we attribute to the important representation of large in size neutral and anionic polyynes in the TMC-1 inventory used during training. Other well-known astrochemical targets, including \ce{CH3NH2}, \ce{CH2NH}, and \ce{CH3O}, not yet detected in TMC-1, but found in other sources, also appear among the candidates.

Finally, the model also identifies some closed shell species with moderate to large electric dipole moments that, in our view, constitute some promising observational targets across different chemical families. These include, \ce{C2NH} ($\overline{\mu}=2.9$ D), \ce{HCCOH} ($\overline{\mu}=1.7$ D), \ce{CH2C(CCH)CH3} ($\overline{\mu}$=0.54 D), \ce{NH2CHS} ($\overline{\mu}$=4.36 D), \ce{HOCCOH} ($\overline{\mu}$= 2.04 D), \ce{CH3OCCH} ($\overline{\mu}$= 1.99 D), \ce{HCCOCCH} ($\overline{\mu}$= 0.77 D) and \ce{CH3C(O)SH} ($\overline{\mu}$=1.84). The electric dipole moments reported in parentheses were obtained after geometry optimization at the $\omega$B97M-D4/def2-TZVPP level using \textsc{ORCA} 6 \citep{mardirossian__2016,neese_software_2025,Weigend2005}. Beyond the hand picked molecules, a full exploration of possible candidates would include, as mentioned above, a more robust clustering procedure of a higher-dimensional UMAP representation, along a high-throughput calculation of the electric dipole moments of the candidates. We defer this more systematic exploration to future work.

Apart from the already discussed regression domain limitation, another limitation for the discovery of new interstellar molecules that is worth discussing in this section is the composition of the prediction subset. While previous works \citep{lee_machine_2021,li_chemical_2026} relied on predicting molecules from databases of commercially available terrestrial compounds, our test set was constructed through an iterative generation of new molecules starting from a predefined molecular inventory, (\texttt{Chaos}). Several regions of chemical space remain inaccessible within our framework. For example, molecules containing elements absent from the TMC-1 inventory, such as Si- or P-bearing species, cannot be generated although an extension of the \texttt{Chaos} protocol to include them would be straightforward. Likewise, pure PAHs are excluded because they cannot be obtained through  substitution. There are also intrinsic chemical limitations in the predicted set. Species such as large radicals, heavy ions, or small anions are poorly represented or absent because the model does not account for their reduced stability or short lifetimes. Large radicals, for instance, recombine efficiently, while small anions rate constant \rev{of formation} is very small \citep{herbst_calculations_2008}. 

\subsection{Test case: Explainability of the recently detected molecules in TMC-1} \label{sec:explainability}

\begin{figure}
    \centering
    \includegraphics[width=\linewidth]{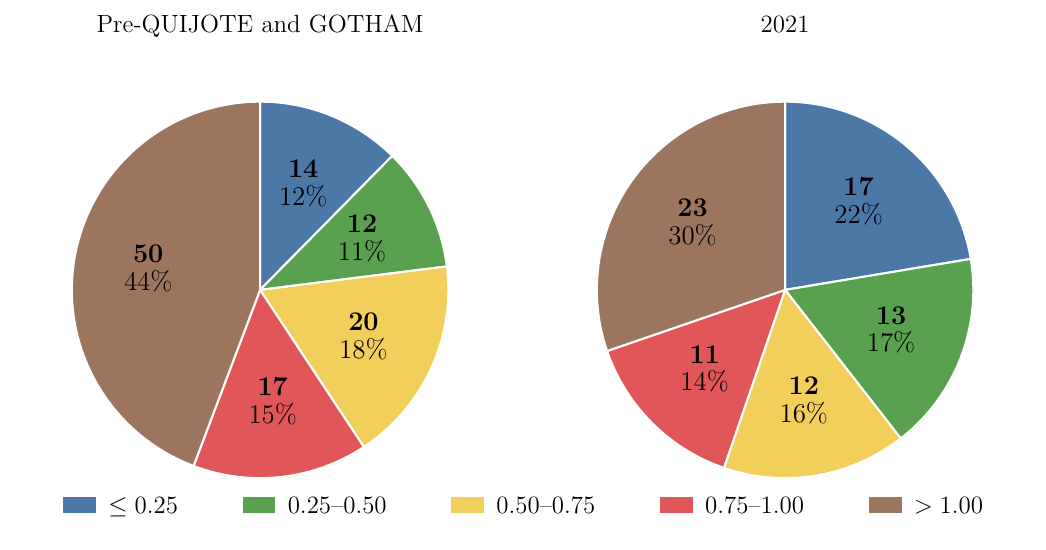}
    \caption{Pie charts showing the percentage of molecules (absolute numbers in parentheses) whose predicted column densities agree with observations within a given error range: brown, more than one order of magnitude; red, 0.75–-1.0 orders of magnitude; yellow, 0.50–-0.75 orders of magnitude; green, 0.25-–0.50 orders of magnitude; and blue, less than 0.25 orders of magnitude. The models were trained using TMC-1 inventories compiled (left) prior to the onset of the \textsc{Quijote} and \textsc{Gotham} surveys, and (right) including all molecules detected in TMC-1 up to 2021. In each case, results are reported only for molecules discovered after the corresponding inventory cutoff dates (approximately 2020 and 2021, respectively)}
    \label{fig:pie}
\end{figure}

The other application we showcase once the model had been trained was to investigate how the recent detections in TMC-1 have reshaped our understanding of the chemistry of this source. As already discussed in the introduction, the unprecedented rate of new molecular detections in TMC-1 has transformed our view of the chemistry not only of this source, but of astrochemistry as a whole. In this work, exploiting the model generalization performance, we make the exercise of determining how the model improves with new molecular detections. In particular, this exercise allows us to assess whether the recently detected species correspond to genuinely novel chemical information or, alternatively, whether their presence could already be inferred from the structure of the previously known molecular inventory. In contrast to the conclusions drawn in Section \ref{sec:inference}, the regression of column densities is fully applicable in this context. The limitation discussed previously arises when the model is applied to arbitrary molecules, for which there is no guarantee that they belong to the same chemical domain as the molecules used during training. Here, however, the evaluation set consists exclusively of molecules that were eventually detected in the interstellar medium. Although these species were unknown at the historical stage represented by a given training set, they remain members of the same underlying population of interstellar molecules

To carry out this analysis, we trained three different models corresponding to distinct stages in the molecular exploration of TMC-1. The first model was trained using the molecular inventory available prior to the \textsc{QUIJOTE} and \textsc{GOTHAM} line surveys, approximately at the time immediately following the detection of benzonitrile (\ce{C6H5CN}; \citealt{mcguire_detection_2018}) in 2018--2020. This initial dataset contains 71 molecules. The second dataset includes all molecules detected up to and including 2021. This period was selected because between 2019 and 2021 alone, 37 new molecules were identified in TMC-1, corresponding to an increase of nearly 50\% in the known inventory. The resulting dataset contains 108 molecules. Finally, the third dataset corresponds to the 184-molecule inventory employed throughout this work \citep[essentially that of][]{agundez_tmc-1_2026}. Each successive dataset fully contains the previous one. The resulting models show performances comparable to those discussed in Section \ref{sec:model_training} using an ensemble of three models. In all cases, we employed the same initialization seed as used for the reference model presented in Section \ref{sec:model_training}.

Once the three models had been trained, we first evaluated their out-of-distribution predictive capabilities on molecules that had not yet been discovered at the corresponding historical stage represented by each dataset. The results are summarized in Figure \ref{fig:pie}. In this figure, molecules predicted with errors below 0.25 orders of magnitude are shown in blue, with progressively different colors indicating larger prediction errors up to deviations larger than one order of magnitude, represented in brown. 

\begin{figure*}
    \centering
    \includegraphics[width=\linewidth]{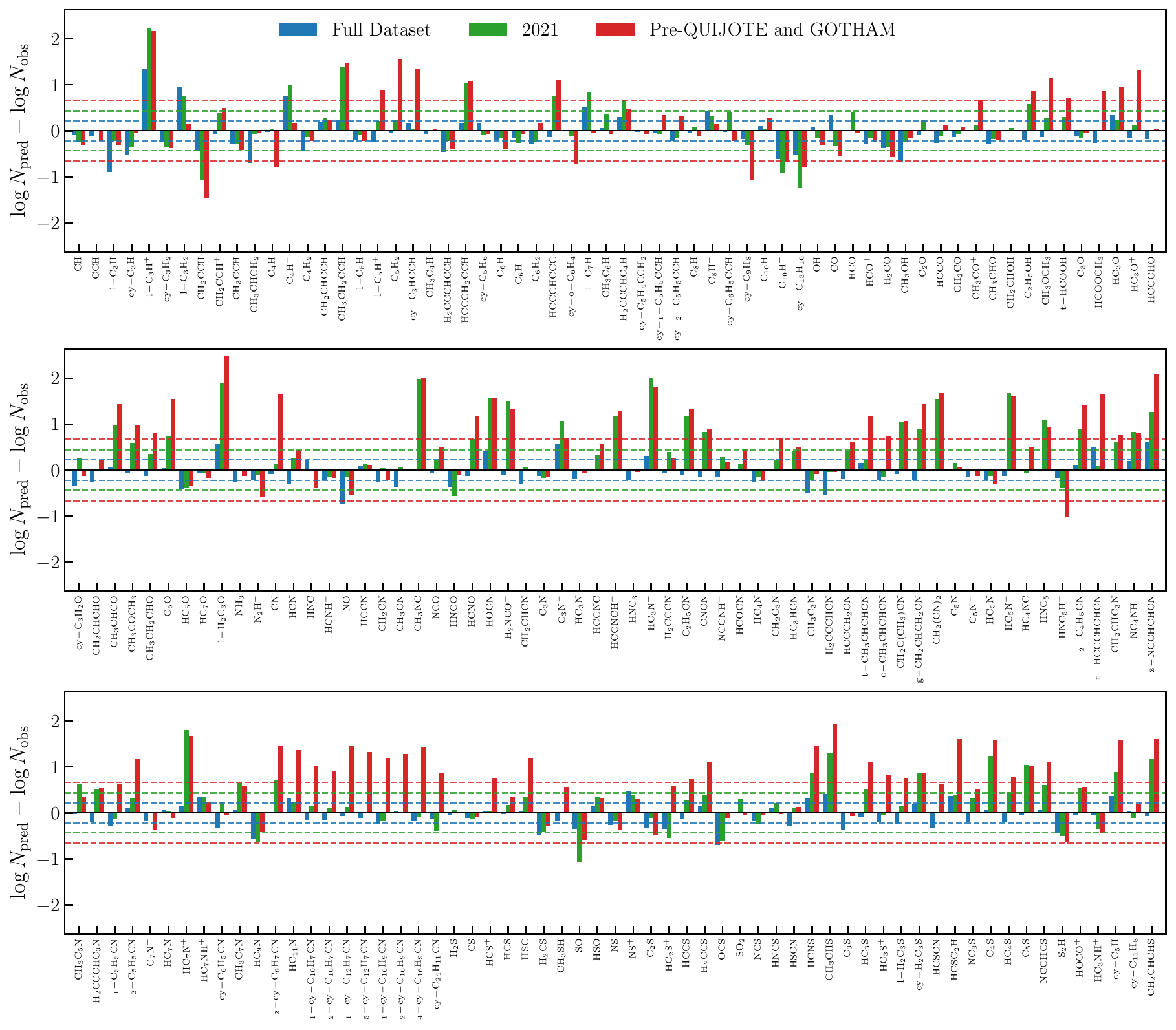}
    \caption{Absolute signed deviations of all the 184 molecules in our dataset trained with different training inventories. In blue, the whole dataset as per January 2026, in green the inventory including all molecules detected in TMC-1 until 2021, and in red the inventory including all molecules before the onset of the \textsc{Quijote} and \textsc{Gotham} surveys (approximately 2020). Horizontal lines represent the MAE for each inventory. The plot is divided in three panels to show the full TMC-1 inventory.}
    \label{fig:bars}
\end{figure*}

It is apparent from Figure \ref{fig:pie} that the model trained on the pre-\textsc{QUIJOTE}/\textsc{GOTHAM} inventory performs significantly worse, with only 12\% of the molecules predicted within 0.25 orders of magnitude of the observed value. In contrast, increasing the training set to the 2021 inventory already results in a significant improvement in performance. This result can be interpreted in two ways. On the one hand, improved performance with a larger training set is trivially expected. On the other hand, and perhaps more interestingly, the substantial improvement obtained after the addition of only 37 newly detected molecules suggests that these species contain genuinely novel chemical information that was not already encoded in the previously known molecular inventory. This finding further supports the idea that recent molecular detections in the ISM are expanding our understanding of interstellar chemistry in directions that could not have been readily anticipated from the pre-\textsc{QUIJOTE}/\textsc{GOTHAM} inventory alone. Moreover, the improvement is not limited to the fraction of molecules predicted within 0.25 orders of magnitude. The relative fraction of molecules with prediction errors between 0.25 and 0.50 orders of magnitude also increases. Nevertheless, at the 2021 stage a substantial fraction of molecules, 30\%, still remain poorly predicted, indicating that the chemical diversity of the interstellar inventory was far from being fully captured by the models available at that time. This same exercise is hardly applicable to the model trained on the full inventory, as all molecules participate in the training process, either in the training or validation sets.  Therefore the model has prior exposure to all molecules in the dataset, leading to performance estimates that are \rev{too} optimistic. However, and naively, we can include a per-molecule analysis of all the detected molecules in TMC-1 for all three datasets, keeping in mind this limitation. The results of this analysis are shown in Figure \ref{fig:bars}, where the absolute error of every molecule detected in TMC-1 excluding conformers and isotopologues is shown. The average absolute deviation for each dataset is represented with a horizontal line of the corresponding color. These horizontal lines are the most informative aspect of the figure, as they contextualize our current understanding of the chemistry of TMC-1. At our current level, the mean absolute deviation from our model is $\sim$0.25. However, and as has been discussed, there are still significant outliers whose presence cannot be predicted a priori, and for which only a posterior analysis of their chemical properties can rationalize the model failure. An open question raised by the strong predictive performance of the model when trained on the current molecular inventory is whether, as the inventory continues to grow, the abundances of newly detected molecules become increasingly predictable from those of the molecules already known. If so, this would suggest that the molecular inventory of TMC-1 might be gradually approaching a regime in which new detections primarily refine, rather than fundamentally reshape, our understanding of the source chemistry.

%-----------------------------------------------------------------

\section{Conclusions, limitations of the model and outlook}

Throughout this work we have been performing a critical evaluation of our model. However, it is relevant to summarize its limitations to also propose avenues of improvement. Possibly, the most pressing limitation of the model is, as anticipated, the inclusion of information on non-detected molecules in the model. Such information would allow us to improve the regression of low abundance molecules, which are likely the most relevant ones for future detections. Besides, the inclusion of non-detected molecules would also likely improve the results in the existing dataset, as new chemical structures would extend its chemical context. A second limitation of the model is the absence of an a posteriori estimation of prediction uncertainties, as it is the case with Bayesian approaches. This latter limitation is of course also linked with the first one, as uncertainty can be used as proxy for the validity of the predictions, and heuristic rules could be developed to flag predictions as potentially unreliable. Finally, our methodology incorporates only a limited information on the tridimensional structure of the molecules or they chemical and energetic properties, as the \texttt{ChemBERTa} embedding is based on the SMILES representation of the molecules, which does not explicitly encode 3D information and the physicochemical descriptors are limited in number and scope.

Despite the limitations, the model presented in this work constitutes a new framework in the derivation of astrochemical intuition or insights from data, following current implementations. The main points derived from our work are the following:

\begin{itemize}
   \item We have finetuned a pretrained chemical language model, \texttt{ChemBERTa}, on molecules derived from the TMC-1 inventory, using heuristic transformations to generate new molecules through a protocol that we named \texttt{Chaos} (Appendix \ref{sec:app:chaos}). The resulting finetuned language model achieves almost equal performance to the original \texttt{ChemBERTa} model in the prediction of molecular column densities in the training dataset, while improving on the metrics in the validation set (Appendix \ref{app:a}). This improvement evinces that the inclusion of characteristic astrochemical molecules allows our model, or in general any similar model, to learn more accurate representations of the chemical space relevant to astrochemistry.
   \item We constructed a multibranch neural network architecture that combines the \texttt{ChemBERTa} fine tuned embedding with physicochemical molecular descriptors. After neural-network hyperparameter optimization, our model achieves metrics comparable with other state of the art models, with an MSE of 0.17, an MAE of 0.28, and an $r^{2}$ of 0.89 in the combined train and validation sets, with a good generalization in the validation set, with an MSE of 0.58, an MAE of 0.59, and an $r^{2}$ of 0.51.
   \item Interestingly, for most of the worst behaving points in the prediction of molecular abundances within our dataset we can rationalize the reason for the failure a posteriori, as they correspond to well molecules with well known peculiar chemical properties, such as \ce{l-C3H+}, \ce{C4H-} or \ce{H2C3O}. 
   \item Profiting from the neural network architecture, we can not only extract predicted column densities for arbitrary molecules, but also a learned vector representation of the astrochemical space, tailored to molecular column densities in an astronomical region. 
   \item Through dimension reduction (UMAP) we identified a geometrical region of related molecules. A posterior projection of a large set of generated molecules ($\sim$50,000) into this region allow us to identify potential candidates for experimental characterization and astronomical search. Among those molecules, we identified typical interstellar molecules such as \ce{H2}, \ce{CH4}, \ce{C2H2}, \ce{C2H4}, \ce{CH3+}, \ce{CO2} or unsubstituted PAHs, but also other molecules with moderate to large electric dipole moments, still undetected in the ISM, including, for example, \ce{C2NH}, \ce{NH2CHS}, or \ce{CH3OCCH}.
   \item We have applied our methodology to the analysis of recent molecular detections in TMC-1 using three models corresponding to different stages in the characterization of its molecular inventory. The model trained on the inventory available before the \textsc{QUIJOTE} and \textsc{GOTHAM} surveys shows a markedly poorer predictive performance for recently detected species than models trained on the 2021 inventory. This finding indicates that the numerous discoveries enabled by these surveys have significantly refined our picture of interstellar chemical complexity and altered the set of molecules that can be reasonably expected to exist in the ISM.
\end{itemize}

Future developments of the model shown in this work will focus on enhancing the ability of the model to capture the three-dimensional structure of molecules. We also plan to investigate rigorous strategies for incorporating non-detections and observational upper limits into the training dataset. Finally, we will explore the application of the model to astronomical sources with markedly different chemical compositions, including oxygen-rich environments, in order to assess its generality and predictive power across diverse astrochemical regimes. More generally, column density guided latent representations such as those developed here could provide a complementary source of astrochemical information for automated spectral-analysis frameworks such as that of \citet{fried_automated_2026}, potentially combining molecular representations tailored to observed column densities with spectroscopic evidence during automated line assignment.

\begin{acknowledgements}
We thank Dr. Markus Dablander for his informative blog post on physicochemical descriptors (\url{https://www.blopig.com/blog/2022/06/how-to-turn-a-molecule-into-a-vector-of-physicochemical-descriptors-using-rdkit/}). \rev{We thank the reviewer of the manuscript for the in-detail assesment of it, that allowed us to improve and much clearly communicate our results}. This work is supported by ERC grant No. 101218790 (\textsc{Isocosmos}) funded by the European Union. Views and opinions expressed are however those of the author(s) only and do not necessarily reflect those of the European Union or the European Research Council Executive Agency. Neither the European Union nor the granting authority can be held responsible for them. GM also acknowledges the support of the grant RYC2022-035442-I funded by MICIU/AEI/10.13039/501100011033 and ESF+ and from PID2024-156686NB-I00 and project 20245AT016 (Proyectos Intramurales CSIC). We also acknowledge funding support from Spanish Ministerio de Ciencia, Innovación y Universidades through grant PID2023-147545NB-I00.

\end{acknowledgements}

\section*{Data availability}

\begin{itemize}
   \item The code used to generate the vector representations of the molecules, list of smiles, and training routines are provided in the following repository \url{https://github.com/gmolpeceres/NN_Astrochem} and list of TMC-1 molecules, smiles and column densities used to train the model.
   \item List of molecules suggested to be abundant in TMC-1 according to the protocol shown in Section \ref{sec:inference} at \url{https://zenodo.org/records/23531603}
   \item The models used in this work are stored at \url{https://zenodo.org/records/23531603}
   \item SQLite library of the neural-network hyperparameter optimization \url{https://zenodo.org/records/23531603}
   \item Any other piece of data or code used in this work is available upon reasonable request to the corresponding author.
\end{itemize}

% WARNING
%-------------------------------------------------------------------
% Please note that we have included the references to the file aa.dem in
% order to compile it, but we ask you to:
%
% - use BibTeX with the regular commands:
%   \bibliographystyle{aa} % style aa.bst
%   \bibliography{Yourfile} % your references Yourfile.bib
%
% - join the .bib files when you upload your source files
%-------------------------------------------------------------------

\bibliographystyle{aa}
\bibliography{bib.bib}

\appendix

\section{Software and acronyms used in this work} \label{sec:acronyms}

\begin{itemize}
   \item Simplified Molecular Input Line Entry System (SMILES): is a text-based, linear, representation of a molecules as a sequence of characters \citep{weininger_smiles_1988}
   \item \texttt{Rdkit}: is an open-source cheminformatics toolkit used to manipulate molecular structures from SMILES and compute physicochemical and topological descriptors for machine-learning analysis \cite{rdkit}.
   \item \texttt{ChemBERTa}: is a transformer-based molecular language model that generates learned molecular embeddings from SMILES representations using self-supervised training, enabling the extraction of chemically informed feature vectors used in the supervised learning tasks \cite{chithrananda_chemberta_2020}. The \texttt{ChemBERTa} model uses \texttt{pytorch} as a default backend \cite{paszke_pytorch_2019}
   \item Domain-Adaptive Pre-Training (DAPT): also referred to as fine-tuning in the main text, is a technique in which a pre-trained transformer model undergoes additional training on domain-specific data, astrochemical molecules in this case, allowing it to specialize in a particular domain while preserving the broad understanding learned from large-scale general model \citep{gururangan_dont_2020}.
   \item \texttt{Keras}: is an open-source deep-learning library that provides a high-level interface for building neural networks, and was used here to implement and optimize the supervised model \cite{chollet2015keras}.
   \item \texttt{Optuna}: library used to identify the set of hyperparameters in the neural network architecture that provides the best agreement with the targets \citep{optuna_2019}.
   \item \texttt{tensorflow}: deep-learning framework used as the backend for \texttt{Keras} in this work \citep{tensorflow2015-whitepaper}.
   \item \texttt{UMAP} is a dimensionality reduction technique that preserves the local and global structure of high-dimensional data, and was used here to reduce the dimensionality of the learned representations for clustering and visualization purposes \citep{mcinnes_umap_2018}
   \item \texttt{scikit-learn}: is an open-source machine-learning library for Python that provides efficient implementations of supervised and unsupervised learning algorithms, data preprocessing tools, model evaluation utilities, and clustering methods, and was used here for various data processing steps (scaling, train/validation splitting) and in the classification of molecules in pre-established clusters in Section \ref{sec:inference} \citep{scikit-learn}.
\end{itemize}

\section{Iterative generation of astrochemical molecules} \label{sec:app:chaos}

We developed a generative scheme in which, starting from an initial astrochemical molecular inventory, isogyric and isocharged, meaning that the resulting molecule preserves the total spin and net charge of the parent molecule, mutations are iteratively applied to each structure. These transformations are  implemented through reaction-based transformations within the \texttt{RDKit} framework. This strategy enables the systematic exploration of chemically plausible regions of molecular space while \rev{grounded in an initial seed, or inventory, that is observationally validated}. We called this scheme \texttt{Chaos} (Chemical Heuristic Astrochemical Organic Synthesis). The input inventory for \texttt{Chaos} is an arbitrary set of molecules, e.g., all interstellar detected molecules. In this work, we used the TMC-1 inventory as the initial seed for the generation of new molecules. From that inventory, and using the \texttt{rdkit.Chem.rdChemReactions} module, we applied a set of 19 different transformations, gathered in Table \ref{tab:chaos_transformations}. The transformations include single, double and triple bond substitutions, which are applied iteratively to each molecule in the inventory. Because several transformations result in the same product, the resulting molecule list is filtered to remove duplicates. Several stopping criteria can be used to control the generation process, like defining a maximum number of heteroatoms, a maximum number of mutations per structure, a maximum of iterations / structures. However, our tests show that an unchecked inventory produces a better training of the finetuned \texttt{ChemBERTa} model. For this work our algorithm we generate a total of 390,899 molecules over the course of three generations, starting from the TMC-1 inventory. 

\begin{table}
\caption{Set of chemical transformations implemented in the \texttt{Chaos} scheme.}
\label{tab:chaos_transformations}
\centering
\begin{tabular}{lc}
\hline\hline
Type & Transformation \\
\hline

\multicolumn{2}{c}{\textbf{Single bond substitutions}} \\
\hline
single & X--H $\rightarrow$ X--Y \\
single & X--Y $\rightarrow$ X--H \\
\multicolumn{2}{l}{\small for all $Y \in \{$CH$_3$, OH, NH$_2$, SH, C$\equiv$N, C$\equiv$CH$\}$} \\

\hline
\multicolumn{2}{c}{\textbf{Double bond substitutions}} \\
\hline
double & X=CH$_2$ $\rightarrow$ X=Y \\
double & X=O $\rightarrow$ X=Y \\
double & X=S $\rightarrow$ X=Y \\
double & X=NH $\rightarrow$ X=Y \\
\multicolumn{2}{l}{\small for all $Y \in \{$\ce{=CH2}, \ce{=NH}, \ce{=O}, \ce{=S}$\}$ with $X \neq Y$} \\

\hline
\multicolumn{2}{c}{\textbf{Triple bond transformations}} \\
\hline
triple & X-C$\equiv$CH $\rightarrow$ X-C$\equiv$N \\
triple & X-C$\equiv$N $\rightarrow$ X-C$\equiv$CH \\
triple & X-C$\equiv$N $\rightarrow$ X-N$\equiv$C \\

\hline
\end{tabular}
\tablefoot{Only molecules composed of atoms in the set \{H, C, N, O, S\} are considered, consistent with the current TMC-1 inventory. Some transformation might be redundant, but the final inventory is filtered to remove duplicates.}
\end{table}

The improvement in the model thanks to the refinement is described in more detail in Appendix \ref{app:a}. The list of molecules is not exclusively used to improve the embedding representation, but a fraction of them is also used as a test set to perform out-of-distribution predictions, as showcased in Section \ref{sec:inference}. This improvement is likely due to the fact that the finetuning process allows the model to learn a chemical representation more closely tied to the specific chemical domain of interstellar molecules, which is not well represented in the original \texttt{ChemBERTa} training set.

\section{Comparison of supervised learning training between \texttt{ChemBERTa} and finetuned model} \label{app:a}

\begin{figure}[ht]
    \centering
    \includegraphics[width=\linewidth]{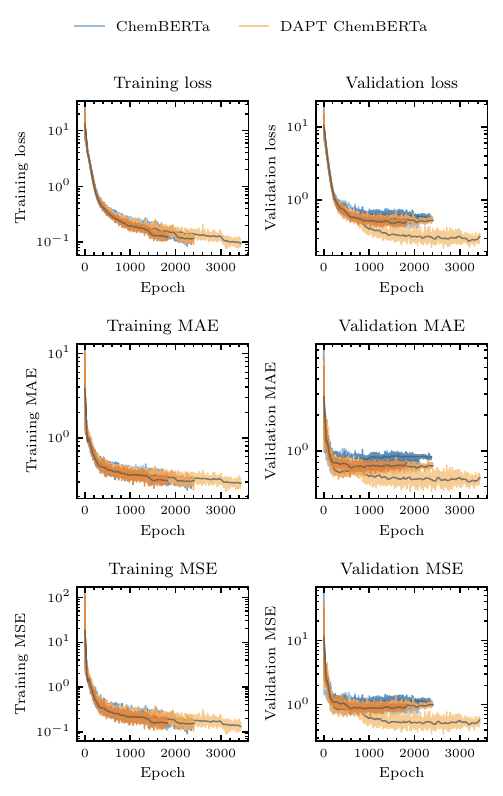}
    \caption{Plots of loss values, MAE and MSE for the training (left panels) and validation (right panels) subsets during the supervised learning training using the original \texttt{ChemBERTa} embeddings (in a range of blue colors) as input for the embedding layer and for the finetuned model (in a range of orange colors) using the molecules generated through the \texttt{Chaos} for steering the molecular representation.  The results are averaged over three different initialization seeds, including different train and validation splits. A Savitzky-Golay filter is applied to each loss curve to ease visualization. }
    \label{fig:chemberta_vs_astrochemberta}
\end{figure}

In the main text we emphasize that the \texttt{ChemBERTa} model used to generate the molecular embeddings was initially finetuned using DAPT to steer the performance of the model in general chemistry of original model to the specific chemical domain of TMC-1 molecules. In this appendix we present a comparison of the performance in the supervised learning training of both models. For that purpose, and using the same architecture, hyperparameters and training/validation splits, we trained generating the molecular embeddings for the training using the original \texttt{ChemBERTa} model and the finetuned one, labelled in this Section DAPT-\texttt{ChemBERTa}, although this notation is not carried over in the main text where \texttt{ChemBERTa} is retained for simplicity. To ensure a fair comparison we track the loss, MAE and MSE during the neural network training for both the training and validation subsets. The results are shown in Figure \ref{fig:chemberta_vs_astrochemberta}. The training curves reveal only marginal differences between the two models, indicating that both embedding representations allow the neural network to fit the training data with comparable accuracy. In contrast, the validation performance is clearly improved in the finetuned model, with consistently lower loss MAE and MSE values for the same splits. The improvement is clear for the validation set, which is the most relevant one to assess the generalization performance of the model. The improvement in the validation performance of the finetuned model is a strong indication that the inclusion of molecules generated through the \texttt{Chaos} protocol allows the model to learn a chemical representation more closely tied to the specific chemical domain of interstellar molecules. Framed in a different direction, this improvement shows that astrochemical molecules are inherently different, at least partially to the molecules used in the original \texttt{ChemBERTa} training, namely the ZINC database of commercially available compounds. Therefore, and optimistically, we can expect that the inclusion of more molecules, not only derived from the TMC-1 inventory, but also from other sources, including molecules for which only upper limits are available, can potentially bring the validation metrics even closer to the training ones. Finally, the almost null improvement in the training performance between models is an indication that the model can easily fit the training data. This behavior is consistent with the relatively large number of trainable parameters in the neural network compared to the limited size of the available training dataset.

\end{document}